\documentclass[twocolumn,floatfix,amssymb,amsmath,secnumarabic,nofootinbib, nobalancelastpage]{revtex4-1}    

\usepackage[pdftex]{graphicx} 
\usepackage{dcolumn}  
\usepackage{bm}       
\usepackage[usenames,dvipsnames]{color}
\definecolor{URLCOL}{rgb}{0,0.52,0.83} 
\definecolor{LINKCOL}{rgb}{0.05,0.5,0} 
\definecolor{CITECOL}{rgb}{0.25,0,0.48} 
\usepackage{epstopdf}
\usepackage[pdftex,bookmarks,breaklinks,bookmarksopen,bookmarksnumbered,colorlinks,
		    linkcolor=LINKCOL,linktocpage,citecolor=CITECOL,urlcolor=URLCOL,
			pdfpagemode=UseOutline,pdftex,pagebackref]{hyperref}
\usepackage{datetime}

\def\preprintlink{ \href{http://dft.uci.edu + PAPER REF}{title of paper} }
\def\preprinttext{~}

\usepackage{fancyhdr}
\makeatletter
\fancypagestyle{titlepage}
{
	\lhead{\textsc{\preprinttext\ ~ \href{http://dft.uci.edu/publications.php}{~}}}
	\chead{}
	\rhead{\preprintlink}
	\lfoot{}
}
\makeatother
\def\preprintlink{ 
	\href{http://dft.uci.edu}
        {
~}
	}

\usepackage{graphicx}
\usepackage{amsmath}
\usepackage{bm}
\usepackage{threeparttable}
\usepackage[normalem]{ulem}
\usepackage{array,multirow}
\usepackage{appendix}
\usepackage{epstopdf}
\usepackage{booktabs}
\usepackage{natbib}
\usepackage{feynmp}
\usepackage{threeparttable}
\usepackage{CJKutf8}
\definecolor{TITLECOL}{rgb}{0.1,0.2,0.7} 
\definecolor{PCOL}{rgb}{0.5,0.06,0.01} 
\definecolor{CHAPCOL}{rgb}{0,0.48,0} 
\definecolor{SECOL}{rgb}{0.1,0.2,0.7} 
\definecolor{CONTENTSCOL}{rgb}{0.1,0.2,0.7} 
\definecolor{SSECOL}{rgb}{0.25,0,0.48} 
\definecolor{SSSECOL}{rgb}{0.2,0.08,0.53} 
\definecolor{SHDCOL}{rgb}{0.4,0,0} 
\definecolor{ITMCOL}{rgb}{0.4,0,0} 
\definecolor{EXCOL}{rgb}{0,0.47,0.01} 
\definecolor{DEFCOL}{rgb}{0,0.42,0.01} 

\def\coltableofcontents{ 
{
\definecolor{SECOL}{rgb}{0.25,0,0.48} 
\definecolor{SSECOL}{rgb}{0.2,0.08,0.53} 
\tableofcontents
}
}

\def\coloredtitle#1{\title{\textcolor{TITLECOL}{#1}}} 

\definecolor{URLCOL}{rgb}{0,0.17,0.43} 
\definecolor{LINKCOL}{rgb}{0.05,0.4,0} 
\definecolor{CITECOL}{rgb}{0.35,0,0.48} 

\definecolor{ngreen}{rgb}{0,0.48,0}

\def\sec#1{\section{\textcolor{SECOL}{#1}}}
\def\ssec#1{\subsection{\textcolor{SSECOL}{#1}}}

\def\sectable#1{
\addcontentsline{toc}{subsection}{~~Table: \textcolor{SSECOL}{#1}}
\begin{table}[h]
\caption{\bf \textcolor{SSECOL}{#1}}
}

\def\bea{\begin{eqnarray}}
\def\eea{\end{eqnarray}}
\def\ben{\begin{equation}}
\def\een{\end{equation}}
\def\benu{\begin{enumerate}}
\def\enu{\end{enumerate}}

\def\bei{\begin{itemize}}
\def\eei{\end{itemize}}
\def\beit{\begin{itemize}}
\def\eit{\end{itemize}}
\def\benu{\begin{enumerate}}
\def\enu{\end{enumerate}}

\def\n{n}

\def\sss{\scriptscriptstyle\rm}

\def\1var{(\bx_1...\bx\N)}

\def\half{\frac{1}{2}}

\def\br{{\bf r}}

\def\bx{{x}}

\def\bj{{\bf j}}

\def\cA{{\cal A}}

\def\s{_{\sss S}}

\def\N{_{\sss N}}

\def\GEA{^{\rm GEA}}

\def\TF{^{\rm TF}}

\def\sph_int{ {\int d^3 r}}

\definecolor{SPECOL}{rgb}{0,0.47,0.01}
\definecolor{QUOCOL}{rgb}{0,0,0.2}
\definecolor{SHDCOLb}{rgb}{0.69,0.4,0.1}

\definecolor{SPEQ}{rgb}{0.01,0.4,0.05} %

\definecolor{SPEQv}{rgb}{0.45,0.05,0.45} %

\definecolor{SPEQb}{rgb}{0.01,0.1,0.65} %

\definecolor{SPEQr}{rgb}{0.57,0.05,0.1} %

\def\sec#1{\section{\textcolor{SECOL}{#1}}}
\def\ssec#1{\subsection{\textcolor{SSECOL}{#1}}}

\def\bay{\begin{array}}
\def\eay{\end{array}}
\def\bit{\begin{itemize}}
\def\beit{\begin{itemize}}
\def\eit{\end{itemize}}

\def\floor{\text{floor} }

\def\WKB{^\text{WKB}}

\def\dd{~ \rotatebox{320}{\hspace{-5pt}\vbox to 5 pt {\hspace{-5pt} \hbox to 5pt {$\cdots$}}}\!\! }

\graphicspath{{Figures/}}
\begin{document}


\sf 
\coloredtitle{Deriving
approximate functionals with asymptotics}
\author{\color{CITECOL} Kieron Burke}
\affiliation{Departments of Physics and Astronomy and of Chemistry, 
University of California, Irvine, CA 92697,  USA}
\date{\today}
\begin{abstract}
Modern density functional approximations achieve moderate accuracy at low
computational cost for many electronic structure calculations.
Some background is given relating the gradient expansion of density functional
theory to the WKB expansion in one dimension, and modern approaches
to asymptotic expansions.
A mathematical framework for analyzing asymptotic behavior for the
sums of energies unites both
corrections to the gradient expansion of DFT and hyperasymptotics of sums.
Simple examples are given for the model problem of orbital-free DFT in one
dimension.  In some cases, errors can be made as small as 10$^{-32}$ Hartree
suggesting that, if these new ingredients can be applied, they might
produce approximate functionals that are much more accurate
than those in current use.
A variation of the Euler-Maclaurin formula generalizes
previous results.
\end{abstract}


\maketitle
\coltableofcontents
\def\floor#1{{\lfloor}#1{\rfloor}}
\def\sm#1{{\langle}#1{\rangle}}
\def\dis{_{disc}}
\newcommand{\Z}{\mathbb{Z}}
\newcommand{\R}{\mathbb{R}}
\def\w{^{(0)}}
\def\w{^{\rm WKB}}
\def\II{^{\rm II}}
\def\hd#1{\noindent{\bf\textcolor{red} {#1:}}}
\def\hb#1{\noindent{\bf\textcolor{blue} {#1:}}}
\def\eps{\epsilon}
\def\ew{\epsilon\w}
\def\ej{\epsilon_j}
\def\upet{^{(\eta)}}
\def\ejeta{\ej\upet}
\def\tjeta{\tj\upet}
\def\bej{{\bar \epsilon}_j}
\def\ewj{\epsilon\w_j}
\def\tj{t_j}
\def\vj{v_j}
\def\F{_{\sss F}}
\def\xt{x_{\sss T}}
\def\sc{^{\rm sc}}
\def\al{\alpha}
\def\ae{\al_e}
\def\bj{\bar j}
\def\bz{\bar\zeta}
\def\eq#1{Eq.\, (\ref{#1})}
\def\cN{{\cal N}}

\sec{Introduction}
\label{intro}

Kohn-Sham density functional theory (KS-DFT) is a very popular electronic
structure method, being used in tens of thousands of papers each year\cite{PGB16}.
However, all such calculations use some approximation to the unknown
exchange-correlation functional of the (spin) densities\cite{KS65}, and most
standard codes allow choices among hundreds (or more) of different
approximations\cite{ED04}, belying claims of a first-principles theory.
There is an {\em exact theory} of DFT exchange-correlation\cite{DG90}, which
is well-developed,
but logically subtle.  This theory shows which properties the exact functional
must have, and which it does not.   Exact conditions are then often
used to determine parameters in approximations\cite{PCVJ92}.  This exact theory
is crucially important in understanding DFT\cite{BW13}, but is not the subject
of this paper.

In elementary quantum mechanics\cite{Gb05}, a standard set of tools is particularly
useful for approximations, such as the variational principle, expansion in 
a basis, and perturbation theory.  These are used extensively in 
traditional {\em ab initio} quantum chemistry\cite{SO82}.  In particular, the
repulsion between electrons is considered as weak, and Hartree-Fock
is the starting point of most methods\cite{F30,H35}.  Most important, in almost
all treatments, a series of equations can be derived of increasing
computational cost to evaluate which (in the case of convergence) will
yield increasingly accurate results.  Similar approaches centering on
the Green's function have been highly successful for calculating responses
of materials, but much less so when used to find ground-state energies\cite{MRC16}.

No such procedure currently exists for density functional theory.
We show here (and in earlier work) that in fact the corresponding
chapter in elementary quantum mechanics is simply that dealing
with semiclassical approximations.  In theoretical chemistry, such
methods were tried long ago in 
electronic stucture (e.g., Refs \cite{M68b,M68c,CKM86}), but are now
more commonly applied
to treating nuclear motion in quantum dynamics\cite{MF97}.
Their exploration for electronic structure withered once modern self-consistent
approximations\cite{S51} could be implemented numerically
with reasonable accuracy.

That this is the unique
perspective from which density functional approximations can
be understood begins with the work of Lieb and
Simon from 1973\cite{LS73,LS77}. Their
work ultimately shows that, for any atom, molecule,
or solid, the relative error in a total energy in a TF calculation must
vanish under a very precise scaling to a high-density, large particle
number limit.  In this limit, the system is weakly correlated, semiclassical,
and mean-field theory dominates\cite{L76,L81}.  This has been argued to
be true also in KS-DFT for the XC energy\cite{C83,EB09,CCKB18}.

The gradient expansion is the starting
place for most modern approximations in DFT (generalized 
gradient expansions\cite{B88,LYP88,PBE96}, and is used in some form in most calculations today.
The first asymptotic correction to the local density approximation for densities that are
slowly varying\cite{LP77,CCKB18} is that of the
gradient expansion.  But a recent paper (hereafter called A\cite{B20}) used an unusual
construction to find the leading correction to the local approximation more generally,
i.e., for finite systems with turning points, for the kinetic energy in one dimension.
It was found that, for such finite systems, the gradient expansion misses a vital contribution,
without which it is much less accurate.

The work of A focuses on the leading corrections to the local approximation.  But these
are just the first corrections in an asymptotic series that, in principle, could be
usefully extended to much higher orders.
Recently, asymptotic methods were developed for sums over eigenvalues for bound
potentials\cite{BB20}, hereafter called B.  
In a simple case, $v(x)=x$ in the half-space $x>0$ in Hartree atomic units,
the sum over the
first 10 eigenvalues was found to be about 81.5 Hartrees, with
an error of about 10$^{-32}$ Hartree.  This extreme accuracy is far beyond any current
computational methods for solving the Schr\"odinger equation.
The way in which this accuracy was achieved employed methods rarely used in 
modern electronic structure calculations, involving hyperasymptotics\cite{BH93}.  Such methods
are difficult to generalize, and often are applied only to very simple, shape-invariant potentials\cite{CKS95},
where specific formulas for the $M$-th order contribution to an asymptotic expansion
can be found explicitly. It would be wonderful if even a tiny fraction of this powerful
methodlogy could be applied to modern electron structure calculations.

The present work is designed as a further step toward this ultimate goal,
as well as a summary of previous work in this direction.
Section \ref{back} summarizes background material from several different
fields.  Section \ref{theory} lays out a general approach to 
summations using the Euler-Maclaurin formula, and shows how
the summation techniques of A and
B are special cases of a this general summation formula.  That
formula yields the key results in both 
A and B, and extends each beyond its original domain of applicability.
I also find a variation that produces the results of A and B simultaneously
and clearly identifies the role of the Maslov index.
I close with a discussion of the relevance of this work to realistic 
electronic structure calculations in Section \ref{DFT}.

\sec{Background}
\label{back}

\ssec{Asymptotics}
\label{asymp}
We begin with some simple points about asymptotic expansions, which we illustrate
using the Airy function\cite{A38,DLMF,VS04}.   Consider an infinite sequence 
of coefficients ${c_n}$, and
the partial sums
\ben
S_M(x) = \sum_{n=0}^M \frac{c_n}{x^n}.
\label{SM}
\een
Consider then $R_M(x)=x^M(S_M(x)-f(x))$.  If
\ben
\lim_{x\to\infty} R_M(x)=0,~~~\lim_{M\to\infty} R_M(x)=\infty
\label{lim}
\een
then $S_M(x)$ is the asymptotic expansion of $f(x)$ as $x\to\infty$, and we write
\ben
f(x)\approx S_\infty (x)
\een
Some important well-known points are that the $c_j$, if they exist, are unique, but infinitely
many different functions have the same asymptotic expansion\cite{C08}.
We shall say that $S_M(x)$ is the $M$-th order asymptotic expansion of $f(x)$.

\begin{figure}[htb]
\includegraphics[scale=.6]{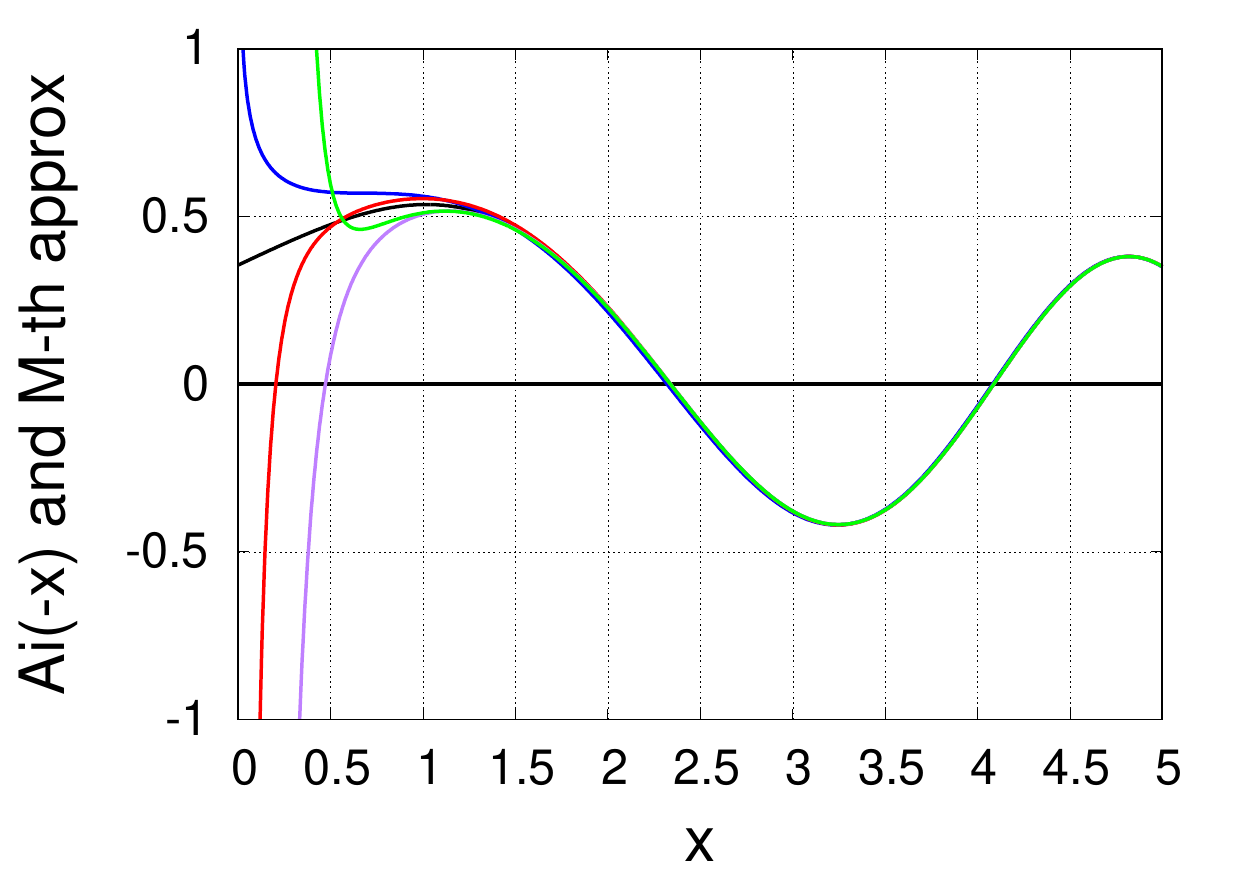}
\caption{$Ai(-x)$ (black) and its asymptotic expansion to zero (blue), first
(red), second (purple), and third (green) orders.}
\label{AiM}
\end{figure}
A simple example is provided by the Airy function of negative argument.   In Fig \ref{AiM}, we plot
this exactly and using its asymptotic expansion of ever increasing order:
\ben
Ai(-x) = \frac{1}{{\sqrt{\pi}}x^{1/4}} 
\Im\{e^{i(z+\pi/4}\, w(z)\}
\label{Aiasy}
\een
where $z=2 x^{3/2}/3$ and
\ben
w(z)= \sum_{j=0}^\infty w_j(z)=1-\frac{5i}{72z}-\frac{385}{10368z^2}+..,
\label{wzasy}
\een
where $w_0=1$, and
\ben
w_{j+1}=-\frac{i}{2z}
\left(j+
\frac{5}{36(j+1)}\right)\ w_j.
\label{wjasy}
\een
From Fig. \ref{AiM} we see that, for $x$ sufficiently large (here about 1.5), the asymptotic expansion
is extremely accurate.
On the other
hand, for $x$ sufficiently small, succesive orders worsen the approximation, and zero-order
is least bad.  Moreover, inbetween, such as at $x=1$, addition of orders at first improves the result
and then worsens it.
\begin{figure}[htb]
\includegraphics[scale=.6]{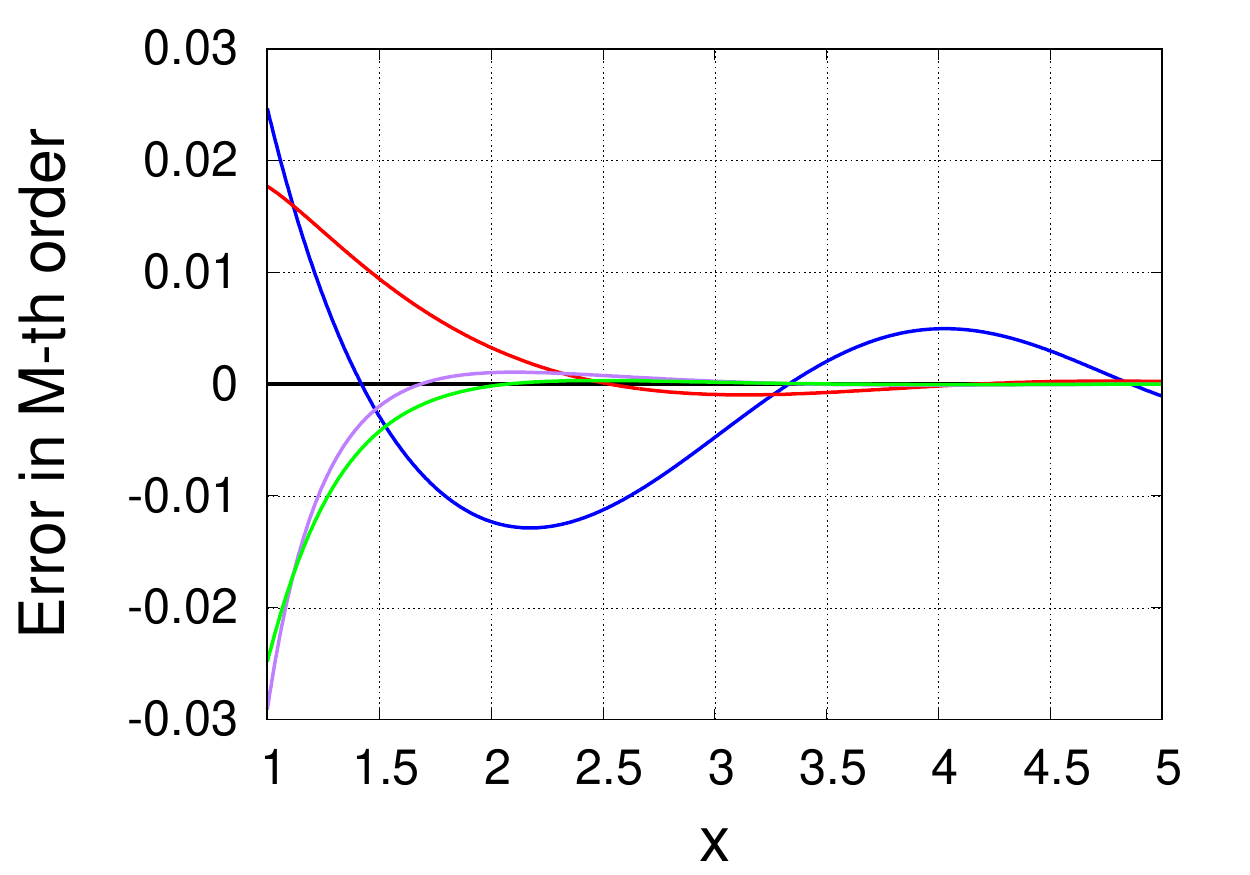}
\caption{Errors at each order, labelling same as Fig \ref{AiM}.}
\label{dAi}
\end{figure}
In Fig. \ref{dAi}, we plot the errors of the successive asymptotic approximatons.  First note
that the scale is 30 times smaller than Fig. \ref{AiM}, and we have begun at $x=1$.  We
see that even at $x=2$, the asymptotic behavior has kicked in, and incredibly tiny errors are
made even for $M=2$.  
To be certain that this is truely an asymptotic expansion, even though the terms in Eq. \ref{wzasy}
appear to be getting smaller, Eq. \ref{wzasy} shows that as $j\to\infty$, $w_j \approx j!/z^j$
which diverges for any value of $z$.

So, suppose we wish to approximate $Ai(-x)$ for all $x$
starting at some finite value, such as $x=1$.
We define $M_o(x)$ as the value of $M$ with the least error, which we refer to as the optimal
truncation.  Then, if we want a `best' approximation to our function by truncating our 
asymptotic expansion, we truncate at $M_o(1)$.  We know that as $x$ increases (at least in the
asymptotic regime), the error of this truncated expansion will reduce.  Thus we expect our
maximum error to be at our lowest $x$, and this truncation will minimize our worst error.
In the top half of Table \ref{T1},
we illustrate this with several orders and several values of $x$.

\begin{table}[htb]
\begin{tabular}{|c|rrrr|c|}
\hline
Order&0&1&2&3&$M_o$\\
\hline
$x$&\multicolumn{4}{c}{Errors}&\\
      0.5&      0.0964&     -0.0069&     -0.3893&      0.1192&1\\
      1.0&      0.0247&      0.0177&     -0.0291&     -0.0248&1\\
      1.5&     -0.0029&      0.0094&     -0.0020&     -0.0042&2\\
      2.0&     -0.0123&      0.0033&      0.0010&     -0.0002&3\\
\hline
$x$&\multicolumn{4}{c}{Additions}&\\
      0.5&      0.5721&     -0.1033&     -0.3824&      0.5085&1\\
      1.0&      0.5602&     -0.0070&     -0.0468&      0.0043&1\\
      1.5&      0.4614&      0.0123&     -0.0114&     -0.0022&2\\
      2.0&      0.2151&      0.0156&     -0.0022&     -0.0012&3\\
\hline
\end{tabular}
\caption{Errors in asymptotic expansion of $Ai(-x)$, and contributions
added at each order.}
\label{T1}
\vskip -0.3cm
\end{table}
But hold on.  We have surely cheated here, because we used our knowledge of the error to
choose where to truncate, which required knowing the function in the first place!  However,
a simple heuristic that usually works is to simply look at the magnitude of the terms that
are being added in each increase in order.  These will typically reduce at first, and then
eventually increase.   The pragmatic optimal truncation procedure is to simply stop when the
next addition is larger in magnitude than the previous one.  We see in Table \ref{T1} that this
indeed corresponds to optimal truncation.

Of greater interest for our purposes will be the asymptotic expansion for the zeroes of $Ai(-x)$,
defined by
\ben
Ai(-a_j)=0,~~~~~j=1,2,3...
\label{ajdef}
\een
in order of increasing magnitude.  Later, we will show that these are the
eigenvalues of a potential.  Each order of truncation of the expansion of 
$Ai(-x)$ in Eq. \ref{Aiasy} implies
an asymptotic expansion of $a_j$ to the same order, yielding
\ben
a_j = y_j^{2/3} \sum_{n=0}^M \frac{T_n}{ y_j^{2n}},~~~~y_j=\frac{3\pi}{2} (j-1/4)
\label{aj}
\een
where the $T_n$ are found and listed in B (appendix B), the first few being
$1,5/48,-5/35,...$.
Because the lowest zero is about 2.34, the asymptotic behavior already dominates for every zero.

So far we have covered basics in most methods books, such as Arfken\cite{A67}.  But now we approach
this from a more modern viewpoint, which holds that often, with the right procedure,
much more useful information can be extracted from such an expansion, especially in cases
that occur in physical problems, i.e., functions that are solutions to relatively simple
differential equations\cite{BH93}.
These methods might generically be called hyperasymptotics\cite{BH99,C08}, and often
begin with the `asymptotics of the asymptotics', i.e., asking what is the behavior of $c_n$
for large $n$ in Eq. \ref{SM}.
  Knowing this, one can use a variety of techniques to approximate the
rest of the sum to infinity, and extract features that are entirely missed in the definition
given above.  However, to take advantage of such techniques, one must be able to write the
expansion to arbitrary order, and then deduce its behavior.

\begin{figure}[htb]
\includegraphics[scale=.6]{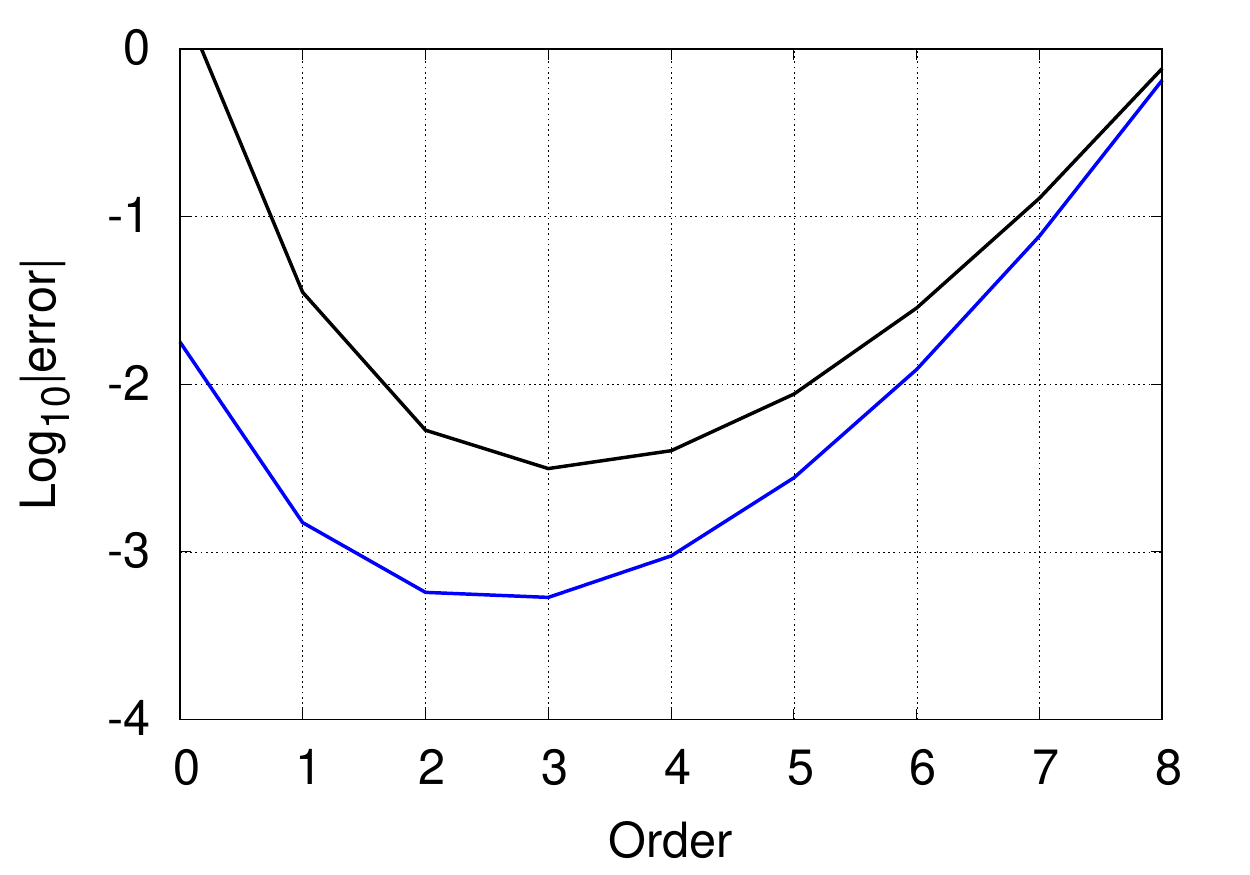}
\caption{Expansion to many orders of $a_1$, the lowest zero of $Ai(-x)$:
the additions (black) and errors (blue).}
\label{A}
\end{figure}

In Fig. \ref{AiM},
we see the first two zeroes, at about 2.34 and 4.09.
In Fig. \ref{A}, we plot both the magnitude of the correction and the
magnitude of the error, on a log (base 10) scale, as a function
of the order of the approximation, $M$, in Eq. \ref{aj}.  We see the generic nature of the asymptotic
expansion.  For small $M$, the additions are quite large. To zero order, the error is of order 0.02.
As more terms are added, the magnitude of the additions becomes smaller, as does the error.
But at $M=4$, the magnitude of the correction is larger than that of $M=3$, so 3 is the optimal
truncation point.  We see that indeed the error also begins to grow.  For large $M$, the
additions become so big that they dominate the error, so the two curves merge.
Thus, with simple optimal truncation, our best possible estimate for the lowest zero is with
$M=3$.

\begin{figure}[htb]
\includegraphics[scale=.6]{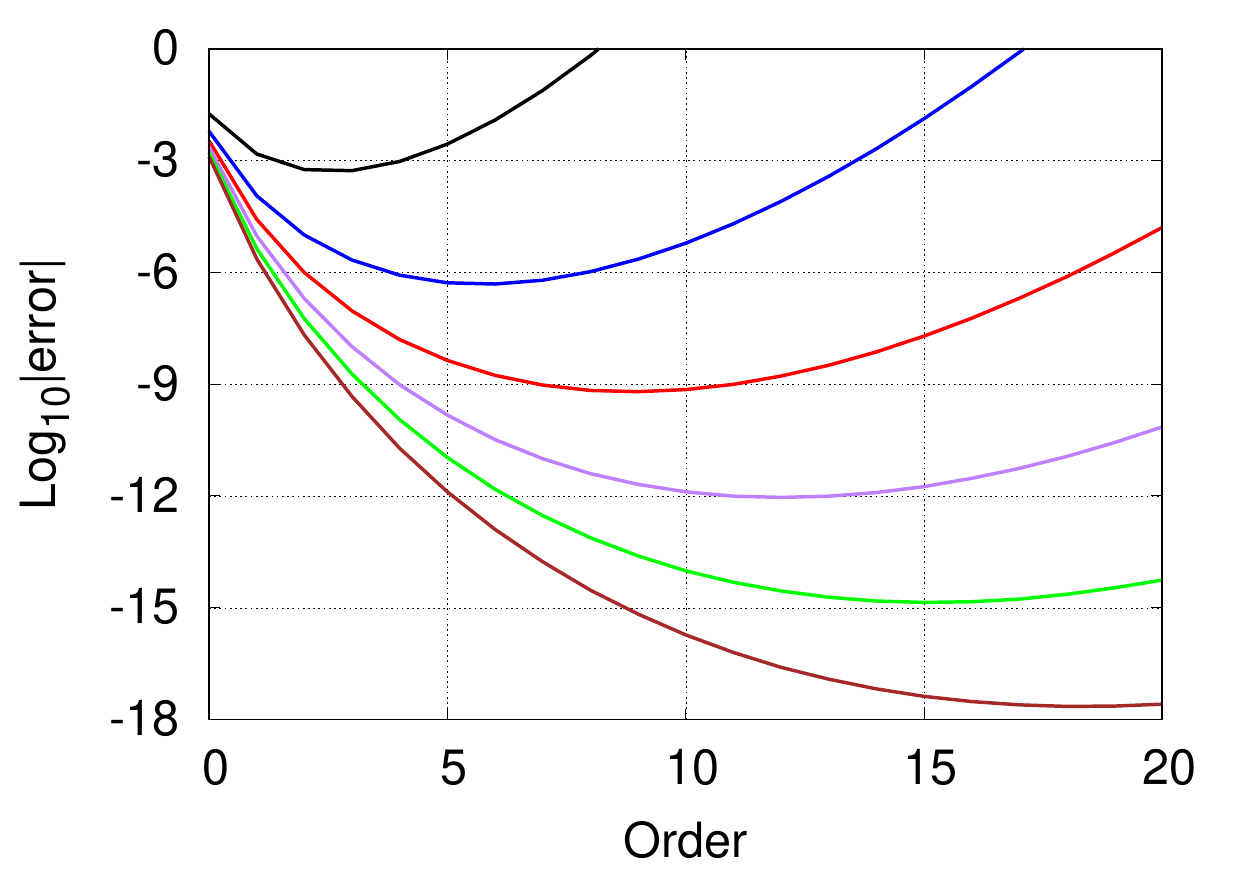}
\caption{Errors of Fig. \ref{A}, but now for first 6 zeroes,
the first being black, the 6th being brown.}
\label{Asy6}
\end{figure}
In Fig. \ref{Asy6}, we show what happens for higher zeroes.  Now the blue curve shows the magnitude 
of the error for the second zero.  Because it is at a higher value, the optimal truncation occurs
at larger $M$, here about 6.  
In fact, the analysis of B shows $M_o(n)\to\floor{\pi*n}$ for the $n$-th zero as $n\to\infty$.
The brown curve is for the sixth level, where the lowest
error is at $M$ about 18, and is of order $10^{-18}$.  This demonstrates the insane levels of
accuracy that can be achieved with very elementary means using asymptotic expansions.

Even the lowest order asymptotic expansion is often rather accurate,
once the asymptotic parameter does not come close to 0.  To write an
approximate formula and apply it to
all zeroes, one should optimally truncate for the lowest level:  All higher levels will then have
lower errors (no lines cross in Fig. \ref{Asy6}).  For any level above the lowest, much greater accuracy
can be achieved by optimal truncation for that level (at a much higher order), but including
those higher orders would be disastrous for the approximation of the lower levels.
For example, for the 6th level, truncation at 18th order yields errors of order $10^{-18}$,
but errors of order $10^{-14}$ for the fifth level, $10^{-10}$ for the 4th, and errors
greater than a Hartree for the lowest two levels.   To get the lowest error for 
every level with a given truncation, Fig. \ref{Asy6} requires truncation at 3rd order.

\ssec{Notation and potentials}
\label{nota}

We choose units with $m=\hbar=1$, so the 1d Schr\"odinger equation is
\ben
\left\{ -\half \frac{d^2}{dx^2} + v(x) \right\} \phi_j(x) = \eps_j\, \phi_j(x)
\een
where $j=1,2,..M$, if only $M$ states are bound.   
We will consider a variety of shapes of potential and boundary conditions.
A hard-wall boundary condition is one where the wavefunction vanishes
identically, and nothing exists beyond the wall.  An asymptotically bound
potential is one where the potential diverges as $x\to\infty$, so that
the system has only discrete states.  Hard walls are a subset of these.
Finally, there is the situation that is closer to realistic, where the
potential is asymptotically free, i.e., tends to a finite constant.
We assume $v(x)$ has a minimum which we choose to be 
at the origin, and set the constant to make $v(0)=0$.  Thus such potentials
tend to $D$ as $x\to\infty$, where $D$ is the well-depth.

\def\PIB{^{\rm PIB}}
Specific examples in this paper include the particle in a box, where
$v=0$ between hard walls at $x=\pm L/2$, with eigenvalues
\ben
\eps\PIB_j = \frac{\pi^2\, j^2}{2L^2}~~~~~(PIB).
\een
and the harmonic oscillator $\omega^2 x^2/2$,
\ben
\eps_j = \omega (j - \half)~~~~~~(HO).
\een
(The unfamiliar minus sign is because the index $j$ begins at 1.)
Another analytically solvable case is the Poschl-Teller (PT) well of depth $D$
\ben
v(x)=D - D/\cosh^2(x),~~\ej = D -(\ae+\half-j)^2/2~~~(PT),
\een
where $\ae={\sqrt{2D+1/4}}$ and $j < \ae+1/2$.  Our last (and most interesting) example
is the linear well $F|x|$, with $F$ a positive constant whose eigenvalues are 
\ben
\eps_{j} = \left(\frac{F^2}{2}\right)^{1/3} d_j~~~~(LW),
\label{epslin}
\een
where $d_{2j+1}$ is the $j$-th zero of $Ai'(-x)$ and $d_{2j}$ is the $j$-th
zero of $Ai(-x)$.

For a symmetric potential $v(x)$, one can always place
a hard wall at the origin.  Then the states of odd parity (even number
with our indexing) become the only eigenstates.  We call these half wells.
For example, for the linear half-well with $F={\sqrt{2}}$, only the even
levels survive, and are given precisely by the zeroes of $Ai(-x)$ shown in Fig. \ref{AiM},
i.e., $d_{2j}=a_j$ of Eq. \ref{aj}.

\ssec{Non-interacting (Kohn-Sham) fermions}
\label{NIFs}

In text books, one usually solves these 1d problems for individual eigenstates.
But we consider these as KS potentials of some many-body problem, presumably with
some approximate XC functional.   As such, we occupy the lowest $N$ levels.
If we keep all spins the same, the total energy is then
\ben
E_N=\sum_{j=1}^N \ej.
\label{EN}
\een
For our simple examples,
\bea
E_N&=&\frac{\pi^2}{6 L^2}\left( N^3+\frac{3}{2}N^2+\half N\right)~~(PIB)\nonumber\\
&=&\omega \frac{N^2}{2}~~~~(HO)\nonumber\\
&=&\frac{\ae}{2}N^2 -\frac{N^3}{6}-\frac{N}{12}~~~(PT)\nonumber\\
&=&\sum_{j=1}^N a_j~~~(LHW).
\label{ENsimp}
\eea
In the last case, there is no simple exact closed form.

The central problem of orbital-free density functional 
theory\cite{T27,F28,M57,T62,LPS84,PY89,DG90,WC00} is to find
sufficiently accurate approximations for $T\s[\n]$, the
kinetic energy of non-interacting electrons as a functional
of their single particle density, $\n(\br)$.  Functional differentiation
and insertion into an Euler equation yields an equation to be
solved directly for the density, avoiding the need 
to solve the KS equations.   Here, the required
level of accuracy is substantially higher than for XC, as the kinetic energy
is comparable to the entire KS energy.   Moreover, since the density of
a given problem will be found by minimizing the energy with the approximate
$T\s[\n]$, the functional derivative must also be sufficiently so that
the approximate density also does not produce an unacceptable error\cite{KSB13}.

In fact, the original Thomas-Fermi (TF) theory\cite{T27,F28} has precisely this form
for the kinetic energy,
but it is not very accurate, its underlying density has many peculiarities,
and it does not even bind
atoms in molecules\cite{T62}.  The form of the TF kinetic energy for spin-unpolarized
electrons in 3d is simply:
\ben
T\s\TF[\n] = \frac{3(3\pi^2)^{2/3}}{10} \int d^3r\, \n^{5/3}(\br).
~~~({\rm 3D,~unpol.})
\een

We will call electrons in a KS potential NIFs, meaning non-interacting fermions.
The effect of the Pauli principle is simply to make them occupy the lowest $N$
orbitals.  Moreover, to avoid keeping track of endless factors of 2\cite{OP79},
we simply choose them all to have
the same spin.  For such spin-polarized NIFS in 1D, the analog of the above is
\ben
T\TF[\n] = \frac{\pi^2}{6} \int_{\infty}^\infty dx\, \n^3(x).
~~~({\rm 1D,~pol.})
\label{TTF1D}
\een
This is of course the local density approximation for the kinetic energy, and is
exact for a fully polarized uniform electron gas.  (As densities scale with
inverse volume, and the kinetic energy operator is a square gradient, in $d$ dimensions
the local density approximation to $T\s$ always has power $\n^{(d+2)/d}$, and
its prefactor is determined by the uniform gas or the large $N$ limit of any
system.)

Thus if one could achieve very high accuracy in an approximate $T\s$
without incurring much computational
cost beyond TF, orbital-free DFT could make solving the KS equations obsolete\cite{WC00}, and
reduce the cost of DFT calculations to that of solving Poisson's equation.
From a regular quantum viewpoint, this is all a very elaborate approach to approximating
the sum in Eq. \ref{EN}.

\ssec{Semiclassical approximations}
\label{semi}

Semiclassical approximations are ubiquitous in physics and chemistry, but are rarely used
directly in electronic structure calculations at present\cite{H18}.  All such expansions involve
powers of $\hbar$ that become relatively accurate in the small $\hbar$ limit. 
Our interest will be in finding eigenstates of the Schr\"odinger equation, specifically in one
dimension.  In this case, the
WKB approximation\cite{W26,K26,B26,D32} is well-known and appears in many introductory text on quantum
mechanics\cite{Gb05}.  The WKB formula for eigenvalues is the implicit formula
\ben
\cA(\eps) = 2\pi (j-\beta/4),~~~~~j=1,2,...,
\label{WKB}
\een
where $\cA$ is the classical action at energy $\eps$
over a complete closed orbit and $\beta$
is the Maslov index\cite{MF01}.   The Maslov index distinguishes between
hard wall reflections and true turning points, i.e., those where the slope of
the potential is finite.  There is no contribution for a hard wall, but for
each true turning point, $\beta$ increases by 1.   For our 1d examples, 
each full orbit yields a contribution equal to double a transit from left to right.
Thus we write
\ben
I\WKB (\eps) = \int_{-\infty}^\infty dx\, p(x,\eps) = j -\nu,
\een
where $p=\Re{\sqrt{2(\eps-v(x))}}$ is the classical momentum at energy $\eps$ in the well,
and $\nu=0$ if there are only hard walls, and increases by $1/4$ for each true
turning point.

We can apply the WKB approximation to each of our wells.  For the PIB, 
$v=0$,
$p={\sqrt{2\eps}}$ and $I\WKB=L p$,  yielding the exact answer as $\nu=0$.
Similarly, for the
HO, and $I\WKB=\eps/\omega$, again yielding the exact answer, as $\mu=1/2$.  The first
can be attributed to the equivalence of semiclassical and exact quantum motion for
a constant potential, the second to the exactness of semiclassical results in harmonic
potentials.  For the PT well,
\ben
I\WKB(\eps) = {\sqrt{2D}}-{\sqrt{2D-\eps}}~~~~(PT),
\een
which recovers the dominant semiclassical approximation, using $\nu=1/2$:
\ben
\eps\WKB (x)={\sqrt{2D}}x - x^2/2~~~~(PT).
\een
Finally, for the linear half-well, with $\nu=1/4$:
\ben
I\WKB(\eps) = \frac{2}{3\pi} \eps^{3/2},~~~
\eps\WKB(x)= \left(\frac{3\pi x}{2}\right)^{2/3}.
\een
In fact, for $F={\sqrt{2}}$, these are precisely the zeroes of the leading order expansion
of $Ai(-x)$ of Sec. \ref{asymp}.  For the linear half-well, the exact expression for $I$ is
\ben
I(\eps)= \frac{z +\Im{(\log w(z))}}{\pi},~~~~z= \frac{2}{3} \eps^{3/2}~~~(LHW),
\label{ILHW}
\een
where
\ben
w (z(x))= {\sqrt{\pi}}x^{1/4}\, e^{-i(z+\pi/4)}\left(Bi(x)+i Ai(-x)\right),
\een
and $Bi(x)$ is the other independent solution of the Airy equation\cite{AS72,DLMF}.
For every well in Sec. \ref{nota},
\ben
\eps_j \to \eps\WKB (j-\nu),~~~~~j\to\infty.
\label{WKBasy}
\een

But the WKB approximation is just the first term in a delicate asymptotic expansion in $\hbar$,
as shown by Dunham\cite{D32}.  We can define an expansion in powers of $\hbar$,
here represented by a dimensionless parameter $\eta$, via
\ben
I_\eta(\eps)=\sum_{k=0}^\infty \eta^{2k}\, I^{(2k)}(\eps),
\een
where $I\WKB$ is just the leading term.  Then
\ben
I_\eta(\eps) = \eta( j-\nu),
\label{Ieta}
\een
determines the eigenvalues implicitly, which can be inverted power by power to yield
an expansion for the energy levels which becomes more accurate as $j$ increases.
This expansion is well-known\cite{BO78}, but is subtle for systems with
turning points.   The naive corrections formally diverge at the turning points,
but these divergences are exactly cancelled by other terms in the wavefunction,
yielding finite contributions in every order.
This is the semiclassical expansion we are interested in, but we wish to find
sums over levels, not individual eigenenergies.

Back in the 1950's and 1960's, there was considerable activity attempting to 
use semiclassical approximations to do electronic structure calculations,
especially by the Miller group\cite{M68b,M68c}.  In fact, Kohn and Sham
developed a remarkably insightful approach\cite{KSb65} just months before
their most popular paper\cite{KS65}, whose ultimate success for numerical
computation overwhelmed interest in semiclassical approaches.
There was much interest in semiclassical methods for more than one dimension
in the area of quantum chaos\cite{M83}.

One can also deduce, e.g., approximate
wavefunctions (see Sec. \ref{unif} below) in the WKB expansion\cite{Gb05}.  Using WKB wavefunctions
to find approximate energies, e.g., by evaluating the Hamiltonian on them,
yields different results\cite{LR91} than those for the eigenvalues, Eq. \ref{WKB}.
All semiclassical methods\cite{BM72} require extreme care in defining
precisely the nature of the expansion and which quantities should be held fixed.

\ssec{Semiclassical limit}
\label{semilim}

We next consider specifically the semiclassical limit for the sum of the energies.  In this
case, one simply integrates the WKB energies over the required number of levels.
For almost any potential,  it can be shown numerous ways that\cite{MP56}
\ben
E_N \approx \int_0^N dx\, \eps\WKB (x) \approx E\TF_N,~~~~\hbar\to 0,
\een
and none of the details of the corrections matter.   The TF approximation can be treated
as a functional of either the density or the potential (see Sec \ref{pot} below) and the
results are the same.
Alternatively, it is a straightforward matter to
extract the kinetic contribution alone\cite{MP56} and find the local density approximation
to the kinetic energy.  Thus, the local approximation (here TF) becomes relatively exact for all
problems in this semiclassical limit.  This is a simple case of the much harder proof
by Lieb and Simon of the same statement for all Coulomb-interacting matter\cite{LS77}.
In the language of Sec \ref{asymp}, the TF theory yields the dominant contribution in an 
asymptotic series for all matter, which implies that its relative error vanishes
in the limit, Eq. \ref{lim}.

\ssec{Gradient expansion in DFT}
\label{grad}

We focus here on the non-interacting kinetic energy, whose gradient expansion was
performed by Kirzhnits\cite{Kc57}, for a slowly-varying gas, using the Wigner-Kirkwood
expansion\cite{W32,K33}.   Ideas of gradient expansions
permeate the HK\cite{HK64} and KS papers\cite{KS65} that created modern DFT,
for both the full functional and
its XC contribution.  The first generalized gradient expansion for correlation was from
Ma and Bruckner\cite{MB68}, from which many modern GGA's are descended.

Here we consider only the non-interating kinetic energy in one dimension.  In that case,
Samaj and Percus did a thorough job\cite{SPb99}, showing how to generate the expansion to arbitrary
order.   We focus on several key points.  First, they expand both the density and the
kinetic energy density as functionals of the {\em potential}.  (This is, after all, how
quantum mechanics normally works.) Given a potential, the expansion for the density is
\ben
n(x)=\frac{k\F(x)}{\pi} \sum_{j=0}^\infty \frac{ a_j(x)}{k\F^{2j}(x)\beta_j},
\label{nv}
\een
where $\beta_j=1-2j$, $k_F = {\sqrt{2(\mu-v(x))}}$ and $\mu$ is
determined by normalization.  The analogous formula for a
kinetic energy density $t(x)$ is found by multiplying by $k_F^2(x)/2$
and replacing $\beta_j$ by $\beta_{j+1}$, where $t(x)$ is a function
whose integral yields $T$.
The coefficients in the expansion are\cite{SPb99}
\ben
a_0=1,~~~a_1=0,~~~a_2=({{k_F}'}^2+k_F k_F'')/4,...
\label{gradcoefs}
\een
Inserting Eq. \ref{nv} into $t(x)$, and expanding in small gradients, yields:
\ben
T\GEA[\n]=\frac{\pi^2}{6}\int dx\ \n^3 (x) - \frac{1}{24}\int dx \left( \frac{dn}{dx}\right)^2+..
\een
This is the exact analog of the usual expansion in 3D\cite{DG90},
except this is the spin-polarized
form, and the coefficient of the von Weisacker contribution is $1/9$ in 3D but $-1/3$ in 1D.
The gradient expansion
is known to 6-th order in 3D\cite{M81} and has been numerically validated under conditions
where gradient expansions apply\cite{YPKZ97}.  It has also been noticed that, for non-analytic
potentials, evaluation of the higher-order terms depends sensitively on 
the boundary conditions\cite{PSHP86}.

\ssec{Potential functionals versus density functionals}
\label{pot}

The creation of modern DFT and the KS equations has clearly been very successful.
However, standard approaches to quantum mechanics yield algorithms that predict,
for example, the energy as a functional of a given potential, $v(x)$ here.  In the
context of KS-DFT, Yang, Ayers, and Wu\cite{YAW04} first clearly showed the relation between potential
functionals and density functionals.   But semiclassical approximations yield
results for a given potential, not density.  Thus Cangi et al\cite{CLEB11,CGB13,CP15} revisited the
entire framework
of density functional theory, from the HK viewpoint, and showed that a logical alternative
was to create a potential functional that also satisfied a minimum principle,
namely
\ben
F_v=F[\n_v],
\een
where $F[\n]$ is the universal part of the energy functional and
$\n_v(\br)$ is the ground-state density of potential $v(\br)$.  Then
\ben
E_0= \min_{v'} \left\{ F_{v'} + \int\, v\, \n_{v'} \right\},
\een
yields the exact ground-state energy and $v=v'$ at the minimum\cite{CGB13}.
Given an expression for $\n_v$,
various strategies can be used to construct a corresponding $F_v$
and so the entire energy can be found.  In the specific case of TF theory, Eq. \ref{TTF1D}
yields exactly the same results for any system whether expressed as a density functional
or a potential functional.
More sophistacted approximations for the
density including higher-order expansions in $\hbar$ (see next section)
are typically not designed
to be variational\cite{GP09}, and minimization might worsen results.  
Such minimizations are not needed if direct application to the external potenial
already yields highly accurate results\cite{CLEB10}.

\ssec{Uniform approximations for the density}
\label{unif}

To find semiclassical expansions for the density as a functional of the potential, one can start
from WKB wavefunctions.  With hard walls, the wavefunctions are simple
\ben
\phi_j(x) \approx \sqrt{\frac{2\omega_j}{p_j(x)}}\sin{\theta_j(x)}\,,
\label{phiWKB}
\een
where $p_j(x)$ is the classical momentum in the $j$-th WKB eigenstate,
$\omega_j$ 
is the frequency of its
orbit, and $\theta_j(x)$ is the phase accumulated from the left wall\cite{ELCB08}.
Using a variation on the standard Euler-Maclaurin formula (Eq. \ref{EMnorm} below)
in asymptotic form, this yields
a uniform approximation to the density.  For any
value of $x$, WKB provides the dominant contribution as $\hbar\to 0$, and its leading corrections
provide the next order in the asymptotic series.  
Early on, it was shown how to extract an accurate approximation to both the density and the kinetic
energy density with hard-wall boundary conditions, by evaluating the next order in the WKB
expansion for the wavefunctions, and summing the result\cite{ELCB08,CLEB10}.   
The leading corrections to the kinetic energy density integrated to yield the
leading correction to the kinetic energy as an expansion in $\hbar$.

However, the WKB wavefunctions are well-known to diverge at a true turning point. 
Langer\cite{La37} found a semiclassical wavefunction that remains uniformly accurate
through the turning point, by replacing $\sin \theta_j$ in Eq. \ref{phiWKB} with
\ben
z_j^{1/4}(x)\, Ai\left[-z_j(x)\right],
\label{phiLang}
\een
where $z_j=  \left[3\theta_j(x)/2\right]^{2/3}$.
Some years later\cite{RLCE15}, and with considerable difficulty\cite{RB18}, it was deduced
how to repeat the same procedure for a well with real turning points, creating a uniform
approximation for the density in a well with turning points, 
i.e., one whose error, relative to the local approximation, vanishes for all $x$
as $\hbar$ vanishes.  Note that the expansion in $\hbar$ is in different
orders depending on how close $x$ is to the turning point.

The resulting approximations are exceedingly accurate for both the density and the kinetic energy
density pointwise, but surprisingly do not yield more accurate energies\cite{RB17}.  When analyzed,
it was found that the expansion in wavefunctions yields energetic corrections of order $\hbar^{1/3}$,
but the leading corrections are of order $\hbar$\cite{RB17}.  This was explained by showing that the 
coefficient of
$\hbar^{1/3}$ vanishes identically!  Thus, this expansion would have to be continued to
two more orders to yield the leading correction to the energy, which might take generations to derive.
Instead, analysis of the simple potentials\cite{RB17} used to test the uniform appoximations showed that
direct sums over eigenvalues can yield the leading corrections to sums over $N$ occupied orbitals,
bypassing (for now) the density and any other real-space quantity entirely.

\sec{Theory}
\label{theory}

\ssec{Summation formulas}
\label{sum}
We begin our theoretical development with an unusual form of
the Euler-Maclarin formula from Hua\cite{H12}
\ben
\sum_{a\leq j \le b} f(j) = \int_a^b dx \left( f(x) + P_1(x) f'(x) \right)
- \left[ P_1 (x) f(x) \right]_a^b
\label{EMHau}
\een
where $a,b$ are real numbers, $f'(x)=df/dx$ must be continuous, and $P_1(x)$ is the
first periodized Bernouilli polynomial\cite{DLMF}.  The periodized Bernouilli polynomials
are 
\ben
P_k(x)= B_k (x-\floor{x}),
\een
where $B_k(x)$ is a Bernouilli polynomial, and 
where $\floor{x}$ is the integer part of $x$.
The Bernouilli polynomials, of order $k$, satisfy many simple conditions, with
the lowest few being
\ben
B_0(x)=1,~~~B_1(x)=x-\half,~~~B_2(x)=x^2-x+\frac{1}{6}.
\een
The famous Bernouilli numbers are then 
\ben
B_k = B_k(1),
\een
which vanish for all odd $k$, except $B_1=1/2$.
Eq. \ref{EMHau} is an unusual form because $a$ and $b$ are continuous.

We next perform the standard trick of repeated integration by parts, leading to
\ben
\sum_{a\leq j \le b} f(j) = \sum_{k=0}^p D_k + R_p,
\label{EEM}
\een
where the end-point contributions are
\ben
D_k=\frac{(-1)^k}{k!} \left[ P_k(x) f^{(k-1)}(x) \right]_a^b,
\een
and the remainder is
\ben
R_p=\frac{(-1)^{p+1}}{p!} \int_a^b dx\, P_p(x)\, f^{(p)}(x).
\een
Eq. \ref{EEM} is true for any $p \geq 1$, so long as the $p$-th derivative of $f(x)$
is continuous.  Here, the term $f^{(-1)}(x)$ is simply the antiderivative, so the
integral is $D_0$.   The $p=1$ case is Eq. \ref{EMHau}.
We call Eq. \ref{EEM} the extended Euler-Maclaurin form, and we use several
variations in what follows.

We note some remarkable features of Eq. \ref{EEM}.  
First, any sufficiently smooth function of $x$ that matches $f_j$ at the integers
yields exactly the same sum, so that all differences in the integral on the right
must be cancelled by the other terms.  
Thus, there are many allowed choices
for $f(x)$ that yield the exact sum.   Adding any sufficiently smooth
$g(x) \sin(\pi x)$ to an acceptable $f$ does not change the sum.
Next, we note that for any range of $a$ and
$b$ between integers, the sum does not change but the integral does, so again such
changes must be absorbed by the remaining terms.  
Lastly, we note that the formula is exact for every $p$.  Choosing a low
$p$ requires less derivatives, but often the remainder term is more difficult
to evaluate.

Of course, there are many different ways to write this formula that are useful in different
contexts.
The special case $a=1_-$
and $b=N_+$ recovers the commonly given form of Euler-Maclaurin,
\bea
\sum_{j=1}^N f_j&=&\int_1^N dx\, f(x) 
+\frac{f(N)+f(1)}{2} + R_p[f]\nonumber\\
&+& \sum_{k=1}^{{\floor{p/2}}} \frac{B_{2k}}{(2k)!}\left(f^{(2k-1)}(N)-f^{(2k-1)}(1)\right),
\label{EMnorm}
\eea
where the remainder term is the same as above.  
The plus sign in the second term on the right occurs because of the discontinuity in 
$P_1(x)$ across an integer, and the vanishing of even derivatives is because all odd
Bernouilli numbers are zero, except $B_1$.
This form is perhaps most familar when
approximating an integral by a sum, but we never use it here.

Next, consider the special case $0 < a < 1$ and $b=N+a$.   Then the sum becomes
specifically that of the first $N$ terms:
\ben
S_N=\sum_{j=1}^N f(j),
\een
while the end contributions simplify to
\ben
D_k(a)= \frac{(-1)^k}{k!}  B_k(a) \left[f^{(k-1)}(x)\right]_a^{N+a}.
\een
We will have use for two special cases.  The first is $a=1/2$, and since\cite{DLMF}
\ben
B_k(1/2)=-(1-2^{n-1}) B_k,
\een
then only even terms contribute to the end-points.
The other case we will use is the limit as $a\to 1$, so that $B_k(a)=B_k$,
and
\ben
D_k(1)= \frac{(-1)^k}{k!} B_k\, \left[f^{(k-1)}(x)\right]_1^{N+1}
\label{Dbmax}
\een
no longer vanishes for $k=1$.
Both these special cases will be of value: The first yields some of the key
the results of A,
the second of B.

\ssec{Sums of eigenvalues}
\label{sumeig}

We first use Eq. \ref{EMHau} to derive a general exact formula for the sum of $N$
energy levels in terms of $\eps(x)$.  From Sec \ref{semi}, we know $\eps_j=\eps(j-\nu)$.
Using this, we find for the sum of the first $N$ energies:
\ben
E_N = \sum_{j=1}^p D_k(a) + R_p(a),
\een
where
\ben
D_k(a)=\frac{(-1)^k}{k!} B_k(a) \left[\eps^{(k-1)}(x)\right]_\al^{N+\al},
\een
and
\ben
R_p(a)=\frac{(-1)^{p+1}}{p!} \int_\al^{N+\al} dx\, P_p(x+\nu)\, \eps^{(p)}(x),
\een
with $\al=a-\nu$.   In the special case $a=\nu$, all integrals and evaluations
run from 0 to $N$.
Finally, for the standard case of two real turning points, insert $\nu=1/2$
in both $D_k(\nu)$ and $R_p(\nu)$ to yield
\ben
E_N=\int_0^N dx\, \eps(x)+\sum_{k=1}^{p} D_k(\half) + R_p(\half).
\een
For the choice $p=1$, because $B_1(1/2)=0$ the end-term vanishes, yielding
the elegant result
\ben
E_N^{2TP} = \int_0^N dx\, \left( \eps(x) + \sm{x} \eps'(x) \right)~~~(\nu=\half,p=1)
\label{E2TP}
\een
where $\sm{x}=P_1(x+1/2)$, and $2TP$ denotes two turning points.
This is recognizable as Eq. 14 of A, but was derived here by more
elementary means.

The analysis of A is confined to $\nu=1/2$ and $p=1$. The current formulas
apply to all possible potentials, i.e., any Maslov index, and allow higher choices
of $p$.  For example, the $p=1$ result for arbitrary $\nu$ yields
\ben
E_N = E_N^{2TP}
+(\half-\nu)\left[\eps(x)\right]_0^N,~~~(p=1)
\label{EN1nu}
\een
i.e., there is a simple correction whenever $\nu$ differs from 1/2.
It is straightforward to check that Eq. \ref{EN1nu} produces the exact sum
when used correctly.  For example, for a half-harmonic oscillator, $\nu=1/4$ and
$\eps(x)= 2\omega x$.  In this case, the 2TP contribution is easy to calculate
as the second term vanishes, due to the constancy of $\eps'$ and the periodicity
of $\sm{x}$, yielding $\omega N^2$.  But there is also a finite addition of
$\eps(N)/4$ to produce  the exact
$E_N = \omega N (N+1/2)$.   Similar corrections are also needed to
recover the exact sum for the particle in a box.
This illustrates the significance of the correct Maslov index in all such calculations.

\ssec{Leading correction to local approximation}
\label{lead}

In this section, we use Eq. \ref{EEM} to
examine just the leading correction to the local approximation.
Because classical action is a monotonically increasing function
of $\eps$ as one climbs up a well, then $\eps(x)$ also grows
monotonically, so its integral grows even more rapidly.  On the
other hand, its derivative will be less rapidly growing, and the
periodic term $\sm{x}$ averages to zero with a constant function.
Thus this term is smaller than the dominant term.

Expanding $I$ in even powers of $\eta$ as in Eq. \ref{Ieta}, we find two leading
corrections to the local approximation to second order:
\ben
\Delta E^{(2)}_N = \int_0^Ndx\, \left( \eps^{(2)}(x) + \sm{x} \frac{d\eps^{(0)}}{dx} \right).
\label{E2}
\een
Thus
there are two corrections:  Those due to the next order in the WKB expansion inside the 
dominant integral
while others are the error made in approximating the sum over WKB eigenvalues by
an integral.  In the case of extended systems where there are no turning points, i.e.,
slowlying varying densities, the spacing between levels goes to zero in the thermodynamic
limit, and the latter correction vanishes.  Thus the gradient expansion of Sec. \ref{grad}
misses such
terms completely.  

In principle, Eq. \ref{EN1nu} also applies to the linear well, but its expansion is more
difficult
than the previous case.  In particular, the asymptotic expansion diverges
at $x=0$, the start of our integral, making it impossible to work with.
We thus use a different version, as developed in B.

\ssec{Hyperasymptotics}
\label{hyper}

We now turn to the work of Ref B.   We see immediately that Eq. \ref{E2TP} is not useful
for asymptotic expansions in powers of $\hbar$, as it includes energies down to zero,
where asymptotic expansions like that of Eq. \ref{SM} diverge.
In fact, we use Eq. \ref{Dbmax}, in which both $a$ and $b$ have been maximized, for a given 
sum from 1 to $N$.  This idea already appeared in the contour chosen in Ref \cite{KSb65},
which circles a pole in the Green's function at $\eps_{N+1}$.  Thus we choose our second
variation to explore asymptotic expansions:
\def\nub{{\bar\nu}}
\ben
D_k(1)=\frac{(-1)^k}{k!} B_k \left[\eps^{(k-1)}(x)\right]_{\nub}^{N+{\nub}},
\een
and
\ben
R_p(1)=\frac{(-1)^{p+1}}{p!} \int_{\nub}^{N+\nub} dx\, P_p(x+\nu)\, \eps^{(p)}(x),
\een
where $\nub=1-\nu$.
The first three $D$'s are:
\ben
\left[E(x)\right]_\nub^{N+\nub}
-\half\left[\eps(x)\right]_\nub^{N+\nub}
+\frac{1}{12}\left[\eps'(x)\right]_\nub^{N+\nub},
\een
where $E(x)$ is the antiderivative of $\eps(x)$.  These forms apply to all wells for any
$p > 0$, but have the advantage of being evaluated at the largest possible energies.

It is trivial to check that these forms yield both the exact results for all
the simple potentials we have encountered so far, for any choice of $p$.
They also recover the leading correction to the semiclassical expansion for the PT
well, producing two corrections, one from the 2nd-order WKB, and the other either
from $D_1$ or $R_1$, just as in Eq. \ref{E2}.

But the real use is in hyperasymptotics, i.e., performing asymptotic expansions to high
orders.  We apply our formulas to the half linear well, so that $\nu=1/4$ and $\nub=3/4$.
We perform the WKB expansion to find an asymptotic series in even powers of $\eta$
for the energies:
\ben
\eps_m(x)=\sum_{p=0}^m\eps^{(2p)}(x).
\een
Inserting this in the summation formula, we chose $p=m$, which guarantees the remainder
term is of order $m+1$ or greater.   The $k$-th end-point term contains orders 
$k-1$ to $m+k-1$ due to the derivatives, but the terms beyond $m$ can be discarded
to find the asymptotic approximant of order $m$.  We can write the result very simply as:
\ben
E_N \approx \sum_{m=0}^M \left( S_m(N+\nub) - S_m(\nub)\right),
\label{ENfromS}
\een
where 
\ben
S_m(x) = \sum_{k=0}^m \frac{B_{2k)}}{(2k)!}
\eps^{(2(m-k),2k-1)}(x)
-\half \eps^{(2m)}(x),
\label{sm2}
\een
where the first superscript indicates the power of expansion in $\eta$ and the second
denotes the number of derivatives taken.
This recovers exactly the expansion in Eq. (6.4) of Ref B.  
The asymptotic expansion is evaluated at $N+1$ rather than at $N$ in the regular 
EM formula.  This confers two distinct advantages: For a given order, our errors are typically
much smaller when the index is increased by 1, and secondly, since the order of optimal truncation
is $\floor{\pi N}$, by evaluating at $N+1$, three additional orders are added to the optimally
truncated series, with their concommitant improvement in accuracy.

We note that one need only evaluate each contribution in Eq. \ref{ENfromS}
at the upper end.  Taking $N=0$
and subtracting then yields $S_N$, guaranteeing correctly its vanishing for $N=0$.
Thus we have recovered the main one-dimensional result of Ref B without need for
(but also missing the elegance of) regularizing sums as $N\to\infty$.
Eq. \ref{ENfromS} can be applied directly to finite wells, such as PT or the truncated linear half-well
of Ref. \cite{BB19}.

\begin{figure}[htb]
\includegraphics[scale=.6]{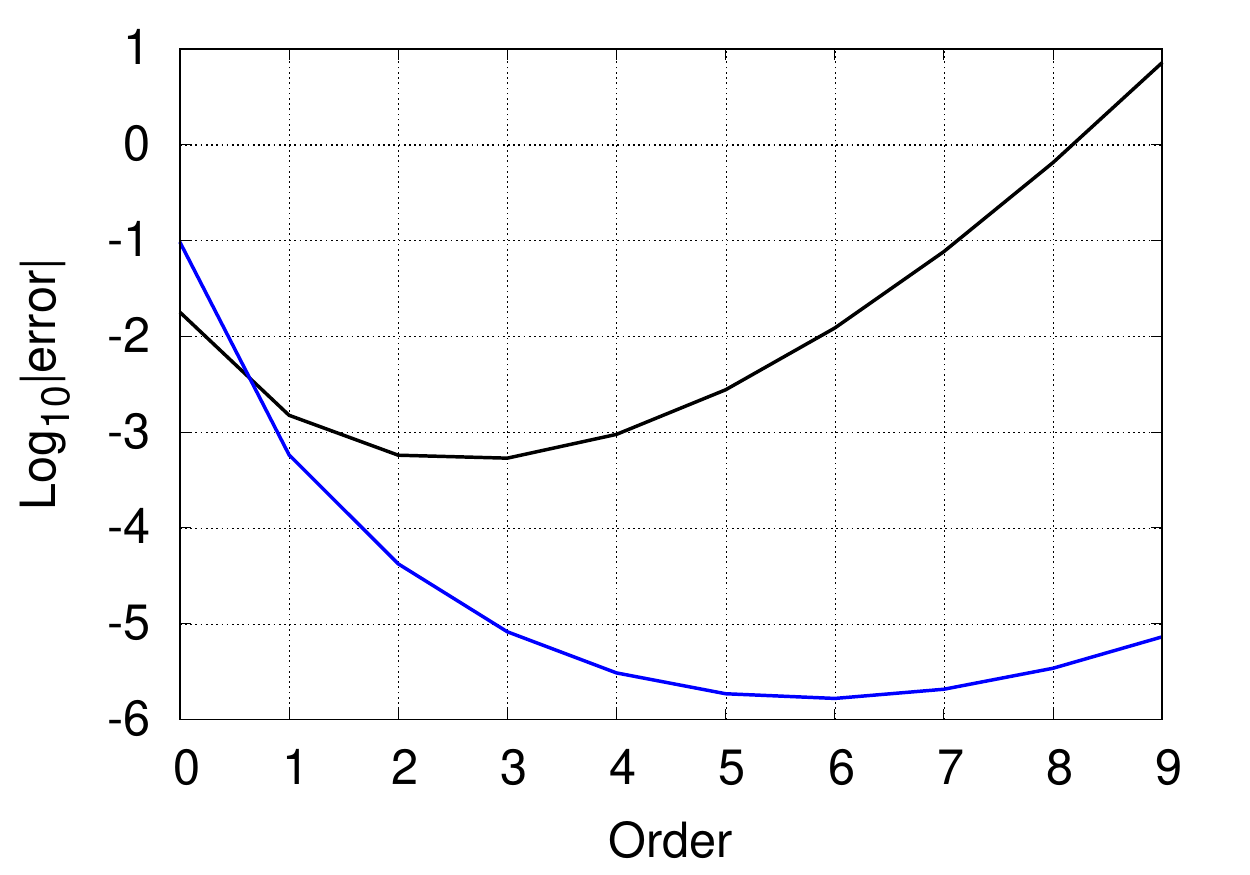}
\caption{Errors of Fig. \ref{A} (black)
and from summation formula, Eq. \ref{Sm} (blue).}
\label{Sum1}
\end{figure}
We show some results from the summation formula for the linear half well
in Fig. \ref{Sum1} for $N=1$,
where $E_1=\eps_1$.  The summation formula is less accurate than the original
formula for $M=0$, but is much more accurate even for $M=3$ (by two orders of
magnitude).  More importantly, its optimal truncation is at 6, producing
almost 3 orders of magnitude in improvement, i.e., going from milliHartrees
to microHartrees errors!

\begin{figure}[htb]
\includegraphics[scale=.6]{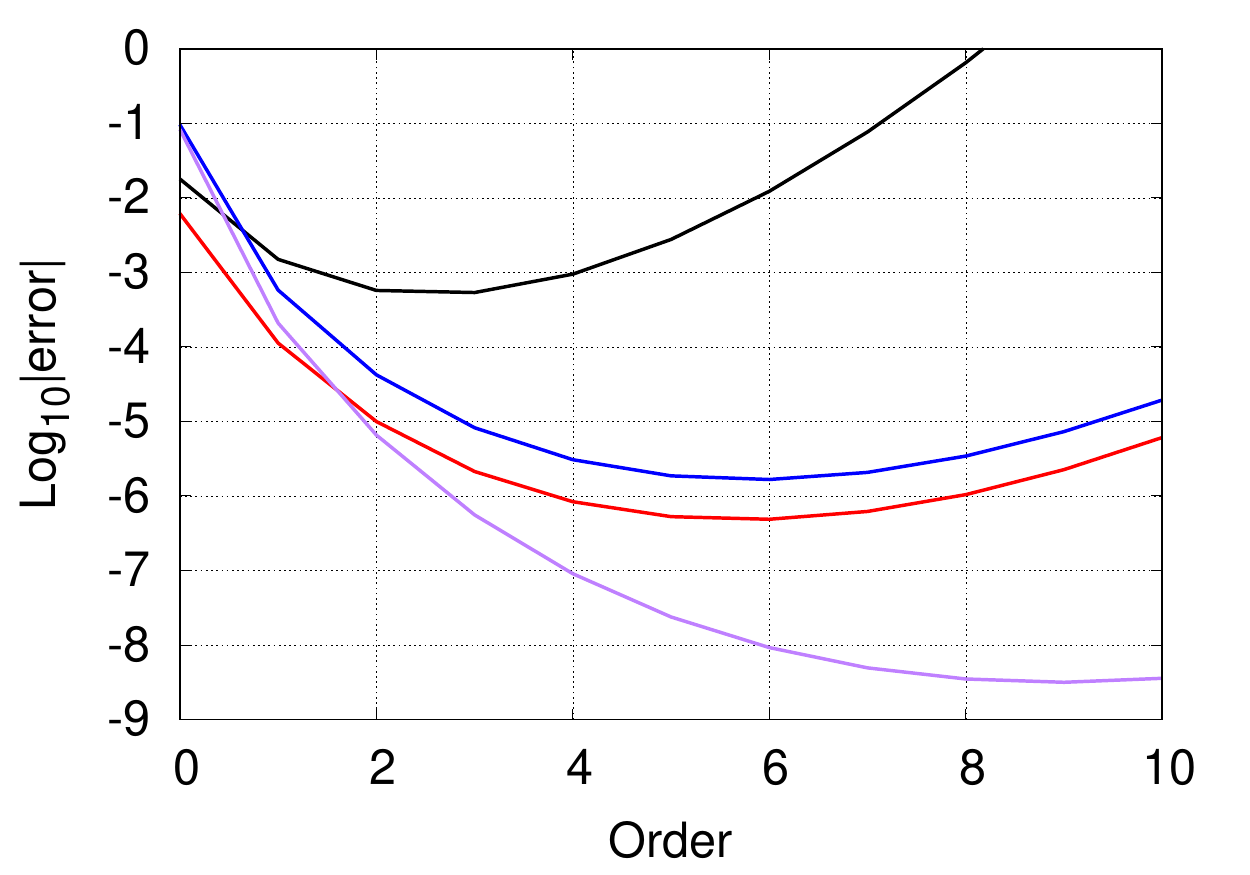}
\caption{Same as Fig. \ref{Sum1}, but adding curves 
for first excited state (red) and the sum of lowest two energies (purple).}
\label{Sum2}
\end{figure}
To see that this is due to our evaluation at $N+1$, in Fig. \ref{Sum2} we add in 
the second eigenvalue and the second sum, $E_2=\eps_1+\eps_2$.
Its error curve is almost identical to that of the
summation formula for the {\em first level}.  Of course, the summation formula
for the 2nd level has leaped ahead again, with errors of nanoHartrees at the
optimal truncation of $M=9$!

\begin{figure}[htb]
\includegraphics[scale=.6]{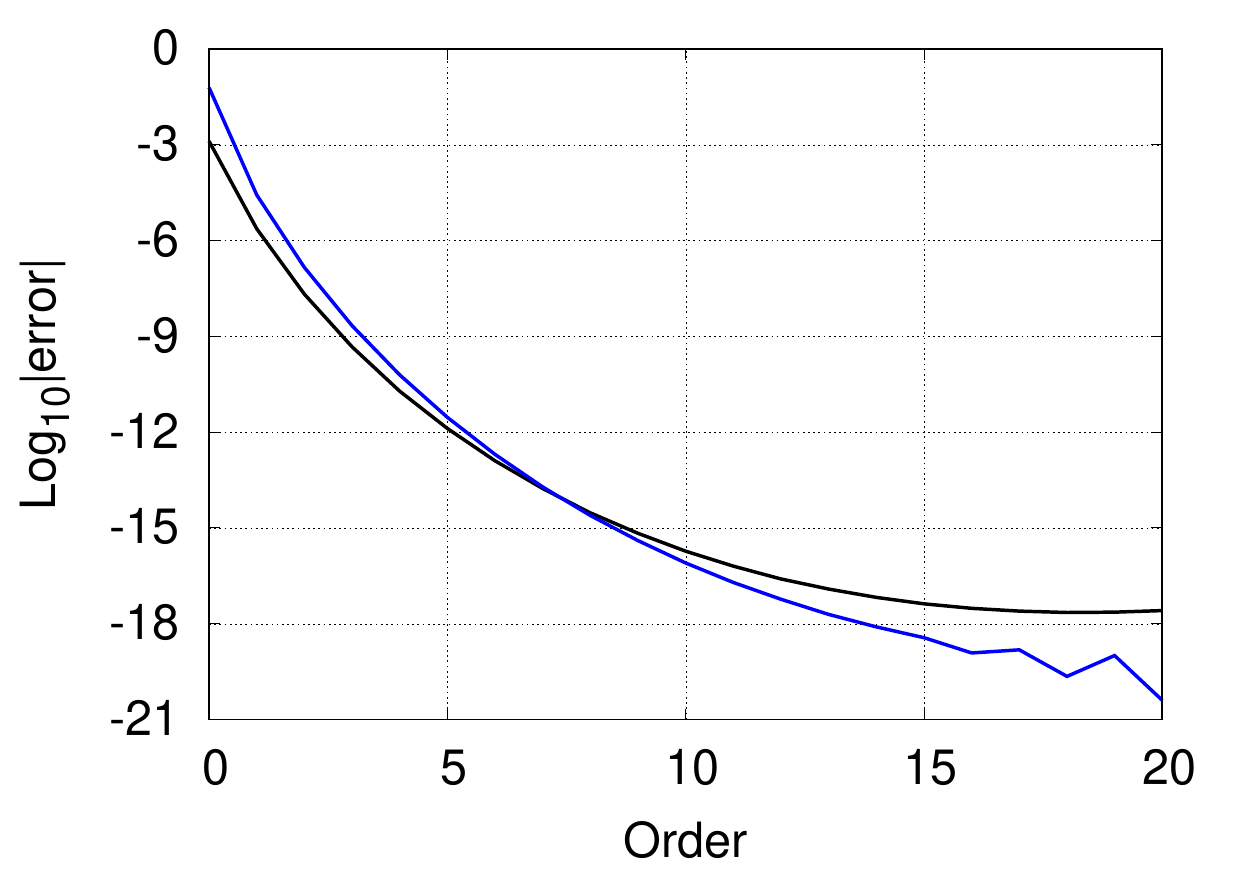}
\caption{Same as Fig. \ref{Sum1}, but for the 6th level and its sum.}
\label{Sum6}
\end{figure}
Finally, we attempt to show the error in the 6th level in Fig. \ref{Sum6}.  The black
line here is for the error in individual level, and so matches the purple line of 
Fig. \ref{Sum1}.  But the blue line is the error in the summation formula, which 
appears to be least at optimal truncation of about $M=21$, where the error is
about 10 zeptoHartrees.  (The noise in the curve is caused by numerical imprecision.)

\ssec{Alternative summation formula}

So far, our analysis has shown that the key formulas of A and B are special cases
of the extended Euler-Maclaurin formula, Eq. \ref{EEM}.   The formulas of A apply only
to two turning points, and cannot be used as a basis for asymptotic expansion, because
the energy function must be evaluated at 0.  The formulas of B require an infinite
set of eigenvalues, but our Eq. \ref{sm2} allows them to be applied to a finite number.
Eq. \ref{sm2} contains the Maslov index explicitly and has no difficulties at the lower-end,
which does not vanish, even in the two turning point case.  

But can we find a single formula that covers all cases?   The primary aim is to
generate an expansion for large $N$, in which one can write exact expressions for
the error.  For any monontonically increasing function of $x$, as our eigenvalues
are defined to be, the large-$N$ limit of the sum is dominated by the integral.
The leading correction is always given by the end-point contribution near $x=N$.
Thus, choosing our upper end-point as $N+1/2$ always eliminates that contribution,
simplifying the result.  Equally, we choose $a=1$ always, so that the lower
energy, even in the presence of two turning points, does not vanish.  This yields
the ungainly but practical
\ben
E_N = \int_a^b dx \eps(x) - \sum_{k=1}^{\floor{p/2}}
\frac{B_{2k}}{(2k)!} (1-\frac{2}{4^k})\, \eps^{(2k-1)}(b) + \Delta_p,
\label{ENfin}
\een
where $b=N+1/2-\nu$, $a=1-\nu$,
and $\Delta_p$ is of order $\eps^{(2p)}(N)$ and is given exactly by
\ben
\Delta_p=-\sum_{k=1}^{p}\frac{B_k}{k!} \eps^{(k-1)}(a) +R_p.
\een
Since the integration interval is no longer an integer, $R_p$ does
not vanish beyond a maximum $p$ for simple powers.
We emphasize that this is an exact formula for all potentials that are
sufficiently smooth (the $p$-th derivative must be
continuous), and can be applied with any $p \geq 1$, and to any boundary 
conditions.  Curiously, almost the same form (but with $a=1/2$) was used in 
Eq. 22 of Ref. \cite{CLEB10} to perform the summation correctly, but without
explanation for why it had this form, or the role of the Maslov index.

For the linear half well, Eq. \ref{ENfin} yields the simple closed-form asymptotic expansion
\ben
E_N \approx \sum_{j=0}^\infty d_j y_N^{5/3-2j},~~~~y_N=\frac{3\pi}{2}\tilde N,
\een
where $\tilde N = N +1/4$, and
\ben
d_j= \frac{2}{3\pi} \frac{T_j} {5/3-2j}-\frac{\pi}{16} T_{j-1} (8/3-2j).
\een
This generates exactly the same asymptotic expansion in $N$ as Eq. 6.4 of B, but in a simpler form
and with terms that differ only by even powers of $z_n$.
The first two terms are:
\ben
E_N\approx \left(\frac{3\pi}{2}\right)^{2/3}
\left(\frac{3}{5}\tilde N^{5/3}-\frac{5+\pi^2}{36\pi^2\tilde N^{1/3}}+...\right).
\label{altasy}
\een
This is identical to, but simpler than, Eq. (6.4) of B.
The GEA of Sec. \ref{grad} includes only contributions from the integral in Eq. \ref{ENfin}.
In the 2nd term, GEA
does not include the $\pi^2$ contribution in the numerator, reducing the overall coefficient
by a factor of about 3, and so misses the correct asymptotic expansion.

\sec{Relation to DFT}
\label{DFT}

\ssec{Error in gradient expansion}
\label{graderr}
While these are impressive ways to sum $N$ eigenvalues, what do they mean for DFT calculations?
We focus on the relation to orbital-free DFT in one dimension (not a very practical
application, admittedly).  

We first consider the direct potential functional form of the gradient expansion, given
in Eq \ref{nv} of Sec \ref{grad}.   All terms can be combined to yield the gradient expansion
for the total energy.  This yields formulas identical to those we find from the
WKB expansion inserted into the integral term, $D_0$, and totally misses the corrections
from the rest of the expansion for any system with discrete levels, as are atoms and molecules.
If this term is included, the results are much more accurate (see Table I of A), 
because the correct asymptotic expansion has been included to the given order.
As shown throughout these works, the sums are much more accurate than the original expansion for the
individual levels.  Without this term, one has only part of the correction, and can
make at best crude guesses (possibly using exact conditions) that yield moderate improvements
at best over the excellent zero-th order contribution.

In Ref A, the correction was first isolated, but only for $\nu=1/2$.  We can now give
the corrections in all cases to every order.  In Eq. \ref{ENfromS}, the gradient expansion
accounts only for those terms of order $m$ in the WKB expansion that occur with the
same order in the summation, i.e., only the $D_0$ contribution to the sum.   Thus
\ben
S^{GEA}_m(x) = E^{(2m)}(x)
-\half \eps^{(2m)}(x).
\label{Sm}
\een
for $m>0$, and the missing terms are
\ben
\Delta S^{GEA}_m(x)=
\sum_{k=1}^m \frac{B_{2k)}}{(2k)!}
D_{2k-1} \eps^{(2(m-k))}(x).
\een
For example, the leading-order missing correction is
\ben
\Delta S^{GEA}_1(x)=
\frac{1}{12} \frac{d \eps^{(0)}}{dx}.
\een
This is the term that was identified in A.  Without it, the 2nd-order gradient
expansion approximation was, at best, an erratic correction to the local approximation.
Including it gave accuracies about 30 times better than the dominant term.
Again, adding the next order led to errors of microHartree order.
Our formulae allow one to extract this missing term to any Maslov index, and so allow it
to be computed, e.g., for the linear half-well.  More importantly, they can be (in principle)
applied to all orders (once the WKB expansion has been performed to a similar order).

\ssec{Understanding aspects of practical KS-DFT}
\label{under}

This 1D world may seem very far from the real world of realistic, practical electronic
structure with Coulomb-repelling electrons being Coulomb-attracted to nuclei,
but many of the difficulties and problems with practical approximate functionals show
up in simpler forms here.  

For example, many semilocal functionals perform worst for one particle.   We see
here that all our results are worst for the lowest level, because the expansion
is asymptotic in the particle number, $N$.  Hence self-interaction error\cite{PZ81}
is a chief source of error in semilocal XC calculations.

Semilocal approximations to XC fail when a bond is stretched, often breaking symmetry at
a Coulson-Fischer point\cite{CF49}.  We see here that, as a bond is stretched, there is a critical
distance in which the well goes from a single well to two.  Beyond that point, 
one can perform the expansion in the separate wells, but it is a different expansion
from the one that applies to a single well.  Thus the asymptotic expansion relevant
at equilibrium becomes irrelevant (and so highly inaccurate) as the well splits in two.
For interacting electrons, this effect is accompanied by a multi-reference character
to the interacting wavefunction.

A third insight, not explored here, is the derivative discontinuities in the energies
as a function of continuous particle number $N$\cite{PPLB82,Pb85}.
The methodology of A demonstrates
this explicitly,
and the present techniques will be expanded to include this in the future.

\ssec{Importance for practical calculations}
\label{import}

Insights from studying these one-dimensional situations have already contributed to
understanding and creating modern functional approximations.  For example, 
the parameter in B88 exchange GGA\cite{B88} was derived in Ref. \cite{EB09}, a
mere 21 years after it was first proposed.   
The derivation yields a value within
10\% of the fitted value of B88 (and is less accurate for real systems).
One of two crucial conditions in constructing PBEsol\cite{PRCV08}, namely the
restoration of the second-order gradient expansion for exchange, came from these
insights.  In fact, the current work may lead to insight into the
second condition, which is the restoration of the LDA surface energy for jellium.
Eq. \ref{altasy} contains a correction missed by the GEA due to the surface
of a linear potential, just the kind of correction being extracted from
the edge electron gas\cite{KMd98,AM05,LMA14}.   
Moreover, Ref. \cite{CCKB18} showed that even the correlation energy of finite systems
finally tends to its LDA value (at least for atoms, but logarithmically slowly).
Several of these asymptotic conditions
for atoms as $N\to\infty$ were built into SCAN\cite{SRP15} and other
approximate functionals\cite{CFLD11}.  Finally, we mention that all chemical
and materials properties depend on energy differences, not total energies.
Ref. \cite{CSPB11} showed that, for atoms with certain plausible assumptions,
the ionization potential is given exactly by KS-LDA calculations in the
asymptotic limit.

It is tantalizing to note that Ref. \cite{EB09} found that the asymptotic
correction to the local density approximation for exchange was almost exactly
double that of the gradient expansion.  This could only be done by numerical
extraction of the coefficient from a sequence of large atom Hartree-Fock calculations.
Eq. \ref{altasy} finds analytically that the correction to the local density approximation
for the total energy of the linear half-well is almost exactly triple that of
the gradient expansion.

\sec{Conclusions}
\label{conc}

I have presented an appropriate mathematical
tool for understanding the successes of modern density functional theory and the
centrality of the local density approximation.   In this framework, 
the continuum limit achieved as $\hbar\to 0$ in a certain, very well-defined sense 
is the reason behind the success of semilocal density approximations.
This framework unites two (apparently) distinct approaches in previous papers, and
generalizes key results from both those works.   More importantly, it shows that,
at least in principle, DFT approximations need not be of low accuracy.  In the simple
case studied here, a well-defined correction has been identified that is missing from
the starting point of most modern approximate schemes, i.e., the gradient expansion,
and its recovery has greatly improved accuracy in model cases.
Further work will follow.

This research was supported by NSF (CHE 1856165).  Kieron Burke thanks Bob Cave,
Attila Cangi, and Raphael Ribeiro
for critical readings of the manuscript.

\bibliographystyle{apsrev}
\bibliography{./Master}

\label{page:end}
\end{document}